\documentstyle[12pt]{article}
\newcommand{\be}{\begin{equation}}
\newcommand{\ee}{\end{equation}}
\newcommand{\ba}{\begin{eqnarray}}
\newcommand{\ea}{\end{eqnarray}}

\begin{document}
\begin{center}
{\bf $SUSY$  INTERTWINING  RELATIONS  OF  THIRD  ORDER  IN  DERIVATIVES}\\
\vspace{0.5cm}
{\large \bf  M.V. Ioffe$^{1,}$\footnote{E-mail: m.ioffe@pobox.spbu.ru},
D.N. Nishnianidze$^{1,2,}$\footnote{E-mail: qutaisi@hotmail.com}  }\\
\vspace{0.2cm}
$^1$ Department of Theoretical Physics, Sankt-Petersburg State University,\\
198504 Sankt-Petersburg, Russia\\
$^2$ Department of Physics, Kutaisi Technical University,\\
4614 Kutaisi, Republic of Georgia\\

\end{center}
\vspace{0.2cm}
\hspace*{0.5in}
%\vspace{1cm}
\hspace*{0.5in}
\begin{minipage}{5.0in}
{\small
The general solution of the intertwining relations between a pair
of Schr\"odinger Hamiltonians by the supercharges
of third order in derivatives is obtained. The solution is expressed
in terms of one arbitrary function. Some properties of the
spectrum of the Hamiltonian are derived, and wave functions for three
energy levels are constructed. This construction can be interpreted as
addition of three new levels to the spectrum of partner potential: a ground state
and a pair of levels between successive
excited states. Possible types of factorization of
the third order supercharges are analysed, the connection with
earlier known results is discussed.
}
\end{minipage}
\vspace*{0.2cm}
\section*{\bf 1. \quad Introduction}
\vspace*{0.5cm}
\hspace*{3ex}

During last two decades the supersymmetrical (SUSY) method \cite{review}
for investigation
of different problems in nonrelativistic Quantum Mechanics (QM) occupied its
stable position among other more conventional approaches. In particular,
it seems to be most adequate for study the properties of isospectrality
(or almost exact isospectrality) of pairs of quantum models and to construct
such pairs ("quantum design"). Correspondingly, the experience of last years
demonstrates that in the SUSY QM algebra the superinvariance of the Superhamiltonian:
\be
[\hat H, \hat Q^{\pm}] = 0
\label{hq}
\ee
represents the main relation. Indeed, most of the known
generalizations and applications of the standard Witten's SUSY QM
\cite{witten} can be formulated as
certain deformations of SUSY algebra, but Eq.(\ref{hq}), as a rule, remains
unchanged (e.g. see \cite{ais}, \cite{acdi}, \cite{ain}).
In terms of the components of the Superhamiltonian $\hat H$ and
supercharges $\hat Q^{\pm}:$
\ba
\hat H =
\left( \begin{array}{cc}\widetilde H&0\\
0& H
\end{array} \right); \qquad
\hat Q^+ =(\hat Q^-)^{\dagger} = \left( \begin{array}{cc}
0&0\\
Q^-&0
\end{array} \right); \quad Q^-=(Q^+)^{\dagger},
\label{definition}
\ea
Eq.(\ref{hq}) takes the form of SUSY intertwining relations:
\be
\widetilde H Q^+ = Q^+ H; \quad Q^- \widetilde H = H Q^-.
\label{intertw}
\ee
Just these relations provide the isospectrality (up to possible
zero modes of $Q^{\pm}$) of the partner Hamiltonians $\widetilde H, H$ and
mutual connection between their wave functions:
\be
\Psi_n(x) = Q^- \widetilde \Psi_n(x);\quad
\widetilde\Psi_n(x) = Q^+ \Psi_n(x).\label{psi}
\ee

The first extensions of {\bf one-dimensional}\footnote{On multi-dimensional
extensions of Witten's SUSY QM see \cite{abi}, \cite{ain} and references therein.}
SUSY QM algebra, which preserve
the intertwining relations (\ref{intertw}), concerned the supercharges
of higher orders in derivatives and were called as the Higher order SUSY
Quantum Mechanics (HSUSY QM), or alternatively, N-fold SUSY QM,
or Non-linear SUSY QM \cite{ais}, \cite{acdi},
\cite{hsusy}, \cite{plyush1},
\cite{andrianov}.

Until now the intertwining relations were completely investigated for
the cases of first and second orders. In particular, two kinds of
second order supercharges were proven \cite{acdi} to exist:
reducible supercharges (are equivalent to two successive first order
supertransformations with real superpotentials) and irreducible supercharges
(are not equivalent).

Much less attention was paid in the literature to the transformations
of third order in derivatives (see the papers \cite{shape},
\cite{kuliy},
\cite{debergh}).
In \cite{shape}
the particular form of third order intertwining relations was
investigated, which leads
to a specific class of shape invariant potentials,
while \cite{debergh}, \cite{kuliy} were
mainly devoted to the construction of potentials with three lowest
energy eigenvalues fixed.
The main goal of this paper is to construct {\bf the general solution}
of intertwining relations
(\ref{intertw}) with supercharges of third order in derivatives.

The paper is organized as follows. The most general solution of
the intertwining relations (\ref{intertw}) with supercharges
of third order in derivatives is derived in Section 2.
From this general solution of intertwining relations some particular
properties of the spectrum are obtained in Section 3,
the explicit expressions
for three wave functions with fixed energy eigenvalues are written.
Thereby one of the partner Hamiltonians, $H,$  can be considered as
quasi-exactly-solvable \cite{turbiner},
\cite{kuliy}. The alternative "quantum design" interpretation is also given here:
starting from $\widetilde H,$ involved in (\ref{intertw}), one can
build its partner $H,$ whose spectrum includes three additional levels (a ground state
ans a pair of levels between excited states).
In Section 4 the variety of possible factorizations of the third order
intertwining operators is investigated. In Section 5 some particular
cases of the general solution are studied, and its classical limit
is discussed.

\vspace*{0.2cm}
\section*{\bf 2. \quad The general solution of third order intertwining}
\vspace*{0.5cm}
\hspace*{3ex} We consider the intertwining relations (\ref{intertw})
on the real axis with the most general supercharges of third order in derivatives:
\be
Q^{+}=(Q^{-})^{\dagger}\equiv \sum_{n=0}^{3}f_{n}(x)\partial^{n};
\quad \partial\equiv\frac{d}{dx};\quad x \in {\cal R}
\label{third}
\ee
where without any loss of generality $ f_{3}(x)=1 $ is chosen.
Thus one obtains a system of nonlinear differential equations for
real functions $ f_{n}(x),\,\,n=0,1,2 $ and real potentials
$ \widetilde V(x),\, V(x):$
\ba
\widetilde V - V &=&2 f^{\prime}_2,\label{1}\\
f_2^2- f_2^{\prime}-2f_1-3V&=&3a,\label{2}\\
2f_2^{\prime}f_1- f_1^{\prime\prime}-2f_0^{\prime}-
3 V^{\prime\prime}-2f_2V^{\prime}&=&0,\label{3}\\
f_0^{\prime\prime}+V^{\prime\prime\prime}+
f_2V^{\prime\prime}+f_1V^{\prime}-2f_0f_2'&=&0,\label{4}
\ea
where $ a $ is an arbitrary real constant, and $ f^{\prime}\equiv df/dx $.
The derivative $f_0^{\prime\prime}$ can be found from (\ref{3}), and after
insertion into (\ref{4}) one obtains:
\ba
(f_1+V)'''-2f_1(f_2'+V)'+2f_2'(V-f_1)'+4f_0f_2'=0.\label{5}
\ea
It is useful to introduce the function
\be
W(x) \equiv V(x)+f_1(x)+a, \label{W}
\ee
in terms of which other unknown functions in (\ref{1}) - (\ref{4}) will be
expressed.
From (\ref{2}) and (\ref{W}) functions $ f_{1}(x) $ and $ V(x) $
can be written via $ W(x) $ and $ f_{2}(x):$
\ba
f_1(x)&=&3W(x)-f_2^2(x) + f_2'(x); \label{fWf}\\
V(x)&=&-2W(x) + f_2^2(x) - f_2'(x) - a. \label{VWf}
\ea
In order to express everything in terms of one function $ W(x) $
it is useful to extract $ f_{0}(x) $ from
(\ref{3}) and (\ref{5}).
Then one obtains the equation:
\ba
(W''+6W^2)'-4W'f_2^2-4f_2'\int f_2W'dx=0, \nonumber
\ea
and after its integration:
\be
W''+6W^2 -4f_2\int f_2W'dx=2c, \quad c=\mbox{const}. \label{aa1}
\ee
Multiplying (\ref{aa1}) by $ W' $ and integrating again,
one derives the relation between $ f_{2} $ and $ W $:
\be
4(\int f_2W'dx)^2=W'^2+4(W^3-cW+d); \quad d=\mbox{const},  \label{6}
\ee
which gives the compact expression:
\be
f_2=\frac{W''+6W^2-2c}{4g}, \label{7}
\ee
where $ g(x) $ is defined\footnote{We suppose that
$ g(x) \neq 0$.} by $ W(x):$
\be
g(x)=\pm\frac{1}{2}\sqrt{W'^2+4(W^3-cW+d)}.\label{8}
\ee
So, both the coefficient functions $ f_{n}(x)$  of
(\ref{third}) for $n=0,1,2 $ and potentials
$ \widetilde V(x),\, V(x) $ are expressed in terms of
an unique arbitrary function
$ W(x),$ thus giving the general solution of SUSY intertwining relations
(\ref{intertw}) of third order in derivatives by Eqs.(\ref{fWf}), (\ref{VWf}),
(\ref{1}), (\ref{3}).

\vspace*{0.2cm}
\section*{\bf 3. \quad Spectrum and wave functions}
\vspace*{0.5cm}
\hspace*{3ex} Results of the previous Section lead to some consequences
for the spectrum and wave functions of
the Hamiltonian $ H.$ By direct (though rather cumbersome)
calculations one can find that the operator $ Q^{-}Q^{+} $
is represented as a third order polynomial of $ H $
with constant coefficients:
\be
Q^-Q^+=(H+a)^3-(H+a)c+d.
\label{qqh}
\ee
Analogously, $ Q^{+}Q^{-} $ is a third order polynomial of
$ \widetilde H $ with the same coefficients, indicating
that the anticommutator of supercharges (\ref{definition})
is a third order
polynomial of the Superhamiltonian, i.e. one obtains
the particular (third order) HSUSY QM \cite{acdi} (see also
\cite{hsusy} - \cite{andrianov}).

It follows from the positivity of $Q^+Q^-$ Eq.(\ref{qqh}) that all
eigenvalues $ E_{n} $ of the Hamiltonian:
\be
H \Psi_{n} (x) = E_{n} \Psi_{n} (x),
\label{eigen}
\ee
satisfy the inequality:
\be
(E_{n} + a)^3 - (E_{n} + a)c + d \geq 0.
\label{inequality}
\ee

We will now restrict ourselves by consideration of wave functions
(\ref{eigen}) which are simultaneously the zero modes of the supercharge:
\be
Q^{+} \Psi_{k} (x) = 0.
\label{zero}
\ee
We denote the energy eigenvalues, which correspond to these $\Psi_k,$
by $\lambda_k;\,\,k=0,1,2.$ It is clear that $ \lambda_{k} $ are the
roots of the third order polynomial:
\be
(\lambda_{k} + a)^3 - (\lambda_{k} + a)c + d = 0.
\label{polynomial}
\ee
We notice that all three roots of (\ref{polynomial})
$ \lambda_{0} \leq \lambda_{1} \leq \lambda_{2} $ are real if the constant
parameters $ d $ and $ c $ satisfy the inequalities:
\be
c > 0;\quad 27d^2 \leq 4c^3 .\label{parameters}
\ee
In this case (\ref{inequality}) implies that $\lambda_0$ necessarily
coincides with the ground state\footnote{We
suppose that the corresponding wave function (see below) is
normalizable.}: $\lambda_0 \equiv E_0,$ while $\lambda_1, \lambda_2$
coincide with two neighbouring excited states: $\lambda_1 \equiv E_j;\,
\lambda_2\equiv E_{j+1}.$ Therefore all other (in general, an arbitrary number)
energy eigenvalues
$ E_{n} \,\,(n\neq 0,j,j+1) $ may belong only to the following intervals:
\be
E_{n} \in \{\bigl(E_{0}, E_{j}\bigr) \cup \bigl(E_{j+1}, \infty \bigr)\}.
\label{interval}
\ee
Thus we deal with the so called
quasi-exactly-solvable \cite{turbiner}
system, i.e. with known positions of the ground state energy and
two additional (not necessarily first excited) energy eigenvalues.
In the case of only one real root $ \lambda_{0} $ of (\ref{polynomial}), it
realizes the ground state energy $\lambda_0\equiv E_0$, and other eigenvalues
(if they exist) are to belong to the semiaxis $ E_{n} > E_{0}. $

These properties of the spectrum of $H$ can also be interpreted from the
"quantum design" point of view. Let us suppose that the spectrum
and eigenfunctions of the {\bf partner} Hamiltonian $\widetilde H$ are known.
Then if $\widetilde H$ is involved in the intertwining
relations (\ref{intertw}), one can construct the Hamiltonian $H$ with
the same (due to (\ref{psi})) energy values  and {\bf three} (in the case of three
roots above)
{\bf additional} levels $E_0, E_j, E_{j+1},$ one of which $E_0$ becomes a
ground state of $H$,
and two other are inserted between a certain pair of {\bf excited} states.
In the case of {\bf one} real root $\lambda_0$
only the ground state $\lambda_0\equiv E_0$ can be included in the spectrum
of $H.$ Of course, saying about
new states $\Psi_n(x);\, n=0, j, j+1,$ we have in mind that their wave functions
are normalizable.
All eigenfunctions of these added states will be constructed below, and their
possible normalizability will be investigated.

By substitution of $ \partial^{2} \equiv \bigl(V(x) - H(x)\bigr) $
the Eq.(\ref{zero}) for $k=0, 1, 2$
is reduced to the linear first order differential equation for
$ \Psi_{n}(x);\, n=0, j, j+1. $ Its general solution reads:
\be
\Psi_{n} (x) = \exp\Biggl( -\int\frac{V' +
(V - E_{n})f_2 + f_0}{V + f_1 - E_{n}}dx\Biggr),
\label{ppsi}
\ee
where $ E_{n};\,n=0, j, j+1  $ is an arbitrary of the roots of (\ref{polynomial}).
By means of (\ref{2}) - (\ref{4}) the nominator in the integrand
can be rewritten as:
$$V' + (V - E_{n})f_2 + f_0=(W - E_{n} - a)f_2 - W'/2 - g(x)$$
to express the wave functions in terms of function $ W(x):$
\be
\Psi_{n}(x) = \mid W(x) - (E_{n} + a) \mid^{1/2}
\exp\int\Biggl(- f_2 + \frac{g(x)}{W - (E_{n} + a)}dx\Biggr),
\label{pppsi}
\ee
where $ f_{2}(x) $ and $ g(x) $ can be found from (\ref{7}), (\ref{8}).
We remark that due to expression (\ref{pppsi}) one can keep the function
$ W(x) $ fixed and study dependence of the wave functions
$ \Psi_{n} (x) $ on their spectral parameters $ E_{n}.$
Vice versa, one can fix the position of
energy level, studying the class of potentials (varying the form
of function $ W(x) $) with the same values of $ E_{n}. $

It is appropriate to remind here that $ W(x) $ in
(\ref{ppsi}), (\ref{pppsi})
is an arbitrary function, which must provide for the
normalizability of eigenfunctions $ \Psi_{n}(x).$ To demonstrate
that this restriction for $ W(x) $ is not too strong, we give
some examples of possible asymptotical behaviour of $ W(x) $ at
infinity and at finite singular point.

1) Let $ W(x) $ has a growing power asymptotics at
$ x\rightarrow\pm\infty .$ Then a negative valued $ g(x) $
in (\ref{8}) $g(x) \sim -W^{3/2}$ and $f_2 \sim \frac{3}{2}W^{1/2}.$
In this case $ \Psi (x)$ is normalizable if $\int W^{1/2}dx \rightarrow
+\infty$ for $x\rightarrow \pm\infty.$

2) Let $ W(x) $ has an asymptotical $x\rightarrow \pm\infty$
behaviour: $W(x)\rightarrow \gamma +\alpha x^{-2},$
where $\gamma(\gamma^2-c)+d=0.$ Then
$g(x)\sim\pm\sqrt{\alpha(3\gamma^2-c)}x^{-1}$
and $f_2(x)\sim2\sqrt{3\gamma^2-c}\alpha^{-1}x.$ Choosing a positive
value for $ g(x) $, one obtains normalizable
$\Psi (x) \sim \exp(-\sqrt{3\gamma^2-c}\alpha^{-1}x^2).$

3) For $d=0$ and $W(x)\rightarrow -\alpha x^{-2},\,\,\alpha > 0$
at $x \rightarrow \pm\infty$ one has $g(x) \sim \pm\sqrt{c\alpha}x^{-1}$
and $f_2(x)\sim\mp\sqrt{c}x/2\sqrt{\alpha}.$ For the negative sign
the wave functions are normalizable:
$\Psi\sim\exp(-\sqrt{c}x^2/4\sqrt{\alpha}).$

4) If $W(x)$ has a singularity at
$ x=0 $: $ W(x) \rightarrow \alpha x^{-2},$
then $g(x)\sim\pm\alpha\sqrt{\alpha+1}x^{-3},$
and $ f_2\sim\pm3\sqrt{\alpha+1}x^{-1}/2.$ For the negative sign
$\Psi\sim \mid x\mid^{\sqrt{\alpha+1}/2-1},$ i.e. $\Psi (x)$
is normalizable at $ x=0 $ for $\alpha>0.$

5) For a pole type singularity of $ W(x)\sim\alpha x^{-1}$
at $x\rightarrow 0$ one can take the negative sign of
$g(x)\sim -\alpha x^{-2}/2$  and $ f_2(x)\sim -x^{-1}$
to obtain normalizable $\Psi(x) .$

In all cases it is necessary to watch nodes of the function $ g(x),$
which generate the singularities of $ f_{2}(x),$ in order to keep
the normalizability of wave functions under the control.

For illustration of the approach we consider a particular example with
$W(x) = \gamma + \frac{\alpha}{1+x^2},$ where, according to item 2) above,
we must choose $\alpha (3\gamma^2-c) > 0$ and $d=\gamma(c-\gamma^2).$
Let us take in addition $\gamma\equiv -\alpha(\alpha +4)/12$ to ensure
that the function $g(x)$ is an entire function:
\ba
g(x)=\frac{\alpha^2 x (x^2 + \frac{\alpha - 2}{\alpha})}{2(1+x^2)^2};\qquad
f_2(x)=\frac{\alpha x}{4} - \frac{\alpha -2}{4x}
-\frac{2x}{1+x^2},
\nonumber
\ea
Then the partner potentials have (up to a common additive constant)
the form:
\ba
V(x) &=& \frac{\alpha^2}{16}x^2 + \frac{(\alpha -2)(\alpha -6)}{16}\frac{1}{x^2};
\label{ppoten1}\\
\widetilde V(x) &=& \frac{\alpha^2}{16}x^2 + \frac{(\alpha^2 -4)}{16}\frac{1}{x^2} +
\frac{4}{1+x^2} - \frac{8}{(1+x^2)^2} + \frac{\alpha}{2}.
\label{ppoten2}
\ea
The singular harmonic oscillator potential (\ref{ppoten1}) on the entire
axis\footnote{Though over this paper we deal with the problems on a whole line, it can be
directly restricted to a half axis $x>0$ by imposing suitable conditions for
behaviour of wave functions at the origin.}
was studied in the literature (e.g. see \cite{singul}). We would not like to make
here the detailed study of this potential our aim, pointing out the paper \cite{dutt}, where
this kind of singular potentials was investigated just in the frameworks of SUSY QM and
unbroken shape invariance. Summarizing \cite{dutt}, the spectrum for the transition (see for
definitions \cite{singular}) "soft" (i.e. for
$-\frac{1}{4} < \frac{(\alpha -2)(\alpha -6)}{16} < \frac{3}{4}$)
potential (\ref{ppoten1}) was shown to consist of two equidistant sequences
of harmonic-oscillator-like levels, and corresponding wave functions were also constructed.

The condition of "softness" of both transition potentials (\ref{ppoten1})
and (\ref{ppoten2}) means
that $\alpha \in (0, 4).$ Thus, the third order intertwining of $V(x)$ and
$\widetilde V(x)$ ensures that
the spectrum and wave functions\footnote{The
nonsingular $(\alpha = 2)$
particular case of (\ref{ppoten2}) was discussed in \cite{kuliy}.} of the potential
(\ref{ppoten2}) coincide (as usual, up to zero modes of $Q^+$) with those of the partner
potential (\ref{ppoten1}).
\vspace*{0.2cm}
\section*{\bf 4. \quad The factorization of intertwining operators}
\vspace*{0.5cm}
\hspace*{3ex} The important question concerns the possible factorization
of the general intertwining operator (\ref{third})
onto the lower order multipliers. We start from the
factorization onto three operators:
\be
Q^{+}\equiv -q_{1}^{+}q_{2}^{+} q_{3}^{+}, \label{factor}
\ee
where $q_{i}^{+}; \,i=1,2,3$ are of the first order in derivatives\footnote{We choose
the signs in operators
$ q^{+} $ the same as in the conventional first order SUSY QM
\cite{review}.}:
\be
q_{i}^{+}\equiv -\partial + W_{i}(x).
\label{ffirst}
\ee
Eq.(\ref{factor}) leads to the following system of nonlinear differential
equations between superpotentials $ W_{i}(x) $ and coefficient functions
$ f_{n}(x):$
\ba
f_2&=&-(W_1+W_2+W_3); \label{21}\\
f_1&=&-(W_{2}+W_{3})^{\prime}+W_{1}(W_{2}+W_{3})+W_{2}W_{3}-W_{3}^{\prime};
\label{22}\\
f_0&=&(W_{2}W_{3})^{\prime} - W_{1}W_{2}W_{3}-W_{3}^{\prime\prime}+
W_{1}W_{3}^{\prime}. \label{23}
\ea
One can take $ W_{2}+W_{3} $ from Eq.(\ref{21}), insert it
into Eq.(\ref{22}) and use the result in Eq.(\ref{23}). In such
a way one obtains the nonlinear second order differential equation
for $ W_{1}(x):$
\be
(W_{1}^{2}-W_{1}^{\prime})^{\prime} -
(W_{1}^{2}-W_{1}^{\prime})(W_{1}+f_{2}) +W_{1}(2f_{2}^{\prime}-f_{1})
-f_{2}^{\prime\prime}+f_{1}^{\prime}-f_{0}=0.
\label{W1}
\ee
Eq.(\ref{W1}) can be linearized by defining $ W_{1}(x) $ as a
logarithmic derivative of a function $ y(x) $:
\be
y^{\prime\prime\prime} - f_{2}y^{\prime\prime} +
(f_{1}-2f_{2}^{\prime})y^{\prime} + (f_{1}^{\prime}-
f_{2}^{\prime\prime}-f_{0})y=0;\quad
W_{1}(x)\equiv -\frac{y^{\prime}(x)}{y(x)}.
\label{log}
\ee
Thus an arbitrary of solutions of (\ref{log}) gives the real superpotential
$ W_{1} $ in terms of real functions $ f_{n}(x).$ Then one can
use (\ref{22}), (\ref{23}) to find also $ W_{2}$ and $W_{3}$ in terms of
$ f_{n}. $

It is necessary to remark that for the real valued solution $ W_{1} $
one can consider {\bf two different opportunities}
for $ W_{2} $ and $ W_{3}:$

1) real valued $ W_{2}, \, W_{3} $; and

2) mutually conjugated complex $ W_{2} = W_{3}^{\star}$ with
{\bf constant} imaginary part\footnote{In this case the third order polynomial
(\ref{polynomial}) has one real and two mutually conjugated roots.}.

The first option can be refered to as realizing a
maximally reducible supercharge (supertransformation). The first order
multipliers $ q_{i}^{+} $ create a chain ("dressing chain"
\cite{veselov}) of Hamiltonians
$ (-\partial^{2} + (W_{i}^{2}\mp W^{\prime} +\epsilon_{i}) $, such that
the neighbouring Hamiltonians (up to a constant)
in the chain are intertwined
by $ q_{i}^{+} .$ This procedure is a particular case of
the familiar construction of reducible Higher order SUSY QM
\cite{ais}, \cite{acdi},
\cite{kuliy} by gluing. In this case
the polynomial (\ref{polynomial}) has three real roots
$ E_{n},\,n=0,1,2.$

The second option
corresponds to partially reducible supercharges, where
$ Q^{+}=q_{1}^{+}M_{23}^{-} $ with irreducible second order operator with
real coefficient functions:
\be
M_{23}^{-}\equiv q_{2}^{+}q_{3}^{+} =
(-\partial + W_{2}(x))(-\partial + W_{2}^{\star}(x)).
\label{M-}
\ee
Though the superpotentials $ W_{2}, W_{3} $ are chosen here to be complex,
the coefficient functions $ f_{n} $ and potentials $ \widetilde V, V $
still are kept real\footnote{The complex valued $ f_{n} $ (with real
$ \widetilde V, V $) would lead to additional intertwining relations of
lower order (see for details the preprint in \cite{shape}
and \cite{andrianov}.}.
Just this kind of intertwining operators was investigated in \cite{ais},
\cite{acdi}.
In this case (\ref{polynomial}) has one real and two
mutually conjugated complex roots $ \lambda_{k} $.

\vspace*{0.2cm}
\section*{\bf 5. \quad Some particular cases of intertwining}
\vspace*{0.2cm}

\subsection*{\bf 5.1. \quad Quasi-exactly-solvable models with
three lowest levels given}
\vspace*{0.2cm}
\hspace*{3ex}
The particular case of the first (maximally reducible) option
above has to be compared with the construction of the paper \cite{kuliy},
where the intertwining operators of third order were useed
to build the quasi-exactly-solvable Hamiltonians,
with three {\bf lowest} values of bound state energy given.
In \cite{kuliy} from the very beginning the intertwining operator had the
factorized form with real superpotentials $ W_{i}(x). $ Those
superpotentials were generated by the {\bf groung state}
wave functions $ \Psi^{(i)}_{0}(x) $ of three consecutive Hamiltonians
$ H^{(i)};\,i=1,2,3, $ included in the chain \cite{acdi} of three
supersymmetric transformations of first order with intermediate gluing (up to
a constant shift):
\be
H^{(i)}\Psi_{0}^{(i)}(x) = E^{(i)}_{0}\Psi_{0}^{(i)}(x);\quad
\Psi_{0}^{(i)}(x)\equiv \exp \Bigl(-\int W_{i}(x)dx \Bigr).
\label{Wpsi}
\ee

It was shown in \cite{kuliy}
that all three superpotentials
$ W_{i}(x) $ and corresponding potentials can be expressed
in terms of one function $ U(x) $
with definite strong limitations in its analytical properties
originated from the restrictions on logarithmic derivative of
normalizable ground state wave functions (\ref{Wpsi}):
$\int W_{i}(x)dx $ has no singularities, and asymptotically
$ \mbox{sign} W_{i}(x)=\pm 1 $ at $ x\rightarrow\pm\infty .$
It can be shown by rather long calculations that the function
$ U(x) $ of \cite{kuliy} and $ W(x) $ in Section 2 of the present
paper coincide up to the constant if the condition (\ref{parameters})
is fulfilled.
Thus the solutions described in \cite{kuliy}
can be obtained as
particular cases of the general construction presented above,
if one will restrict himself by suitable strong conditions for analytical
and topological properties of functions and for relations
between constants. In the general construction of the present paper
both the singularities and the complexity of functions $ W_{i} $ are allowed
if they preserve potentials $ \widetilde V, V $ to be real and not
too singular.

We remark that each of two kinds of factorization of $ Q^{+} $
depends on one arbitrary (up to some discrete restrictions) function.
It is obvious that the alternative to option 2) partially
reducible factorization of $ Q^{+} $ exists, if one chooses $ W_{3} $
real and
$ W_{1}=W_{2}^{\star} $ with constant imaginary part. The third
similar opportunity with $ W_{2} $ real and $ W_{1}=W_{3}^{\star} $ is
equivalent to the previous ones due to commutation
$ [q_{1}^{+}, q_{2}^{+}] = [q_{3}^{+}, q_{2}^{+}] = 0. $

\vspace*{0.2cm}
\subsection*{\bf 5.2. \quad Models with shape invariance of
third order}
\vspace*{0.2cm}
\hspace*{3ex} We will now consider the particular case of
intertwining relations (\ref{intertw}) of third order where
$ \widetilde H $ and $ H $ {\bf coincide up to a constant}
$ 2\lambda :$
\be
H Q^+ = Q^+ (H + 2\lambda). \label{d1}
\ee
This (shape invariant) sort of intertwining appeared in \cite{shape}
in a natural way when a pair of Hamiltonians
$ \widetilde H, H $ participated simultaneously in two intertwining
relations - of first (by operators $ q^{+})$
and second (by operators $ M^{-})$ order in derivatives.
\be
q^{+} = -\partial + \omega (x));\quad
M^{-} = \partial^2 + 2\phi(x)\partial + b(x) + 2\phi^{\prime}(x)),
\label{qM}
\ee
where (see \cite{shape}, \cite{acdi}
for details)
\ba
b(x) &=& -\phi^{\prime}(x) + \phi^{2}(x) -
\frac{\phi^{\prime\prime}(x)}{2\phi (x)} +
\frac{\phi^{\prime 2}(x)}{4\phi^{2} (x)} +
\frac{\beta}{4\phi^{2} (x)};\quad \beta > 0;
\label{b}\\
\phi^{\prime\prime}(x) &=& \frac{(\phi^{\prime}(x))^2}{2\phi (x)} +
6 \phi^{3}(x) + 8\lambda x \phi^{2} (x) +
2(\lambda^{2}x^{2}-(\lambda + \alpha))\phi (x)+ \frac{\beta}{2\phi (x)}
\label{phi}
\ea

The specific shape invariance \cite{genden}
property of the
Hamiltonian $H,$ realized by (\ref{d1}), allowed to find
its spectrum algebraically \cite{shape}
by analysis of zero modes of $ Q^{+} $ and $ Q^{-}:$
\be
Q^{+}\equiv q^{+}M^{-};\quad Q^{-} = M^{+}q^{-}.
\label{QqM}
\ee

Comparing coefficients of $ \partial^{k} $ in (\ref{QqM}), one obtains
the relations:
\ba
f_{2}(x) = 2\phi (x) - \omega (x);
\nonumber\\
f_{1}(x) = 4\phi^{\prime}(x) - 2\omega(x)\phi (x) + b(x).
\label{comparef1}
\ea
The function $ f_{2}(x) $ is fixed by (\ref{1}) and (\ref{d1}):
\ba
f_{2}(x) = -\lambda x; \quad \omega (x) = 2\phi (x) + \lambda x.
\nonumber
\ea
It follows from Eqs.(\ref{fWf}), (\ref{aa1}) and (\ref{comparef1}) that
\ba
W(x) = \phi^{\prime}(x) -2\phi^{2}(x) -2\lambda x\phi (x) +
\frac{2\lambda + \alpha}{3}
\nonumber
\ea
and, due to (\ref{aa1}), has to satisfy nonlinear differential
equation:
\be
\Biggl(\frac{W^{\prime\prime}(x)+6W^{2}(x)-2c}{x}\Biggr)^{\prime}
= 4\lambda^{2} x W^{\prime}(x).
\label{equa}
\ee
Indeed, it can be shown by direct calculation (using (\ref{phi})) that
(\ref{equa}) is satisfied for
$ c=\frac{(2\lambda +\alpha)^{2}}{3}-\beta .$ Thus the shape invariant
case (\ref{d1}) is included in the general solution of Section~2.

\vspace*{0.2cm}
\subsection*{\bf 5.3. \quad Models with intertwining of
first and third orders}
\vspace*{0.2cm}
\hspace*{3ex}
The third order intertwining of the previous Subsection was based
on the {\bf double intertwining} of the same pair of Hamiltonians
$ \widetilde H, H $ - by operators $ q^{+} $ of the first
and  $ M^{-}$ of the second orders. Now we will consider the
double intertwining by operators $ q^{+} $ of first order and
$ Q^{+} $ of third order in derivatives. The first intertwining gives:
\ba
\tilde H = q^{+}q^{-} = -\partial^{2} +
\omega^{2}(x) - \omega^{\prime}(x) + \epsilon;\quad
H = -\partial^{2} + \omega^{2}(x) + \omega^{\prime}(x) + \epsilon;\,\,\,
\epsilon = \mbox{const},
\nonumber
\ea
while the second one (see (\ref{1}), (\ref{VWf})):
\ba
V(x) = -2W(x) + f_{2}^{2}(x) - f_{2}^{\prime}(x) - a = \widetilde V(x)
- 2f_{2}^{\prime}(x).
\nonumber
\ea
Thus,
\ba
\omega (x) = -f_{2}(x) - \gamma;\,\,\gamma = \mbox{const};
\nonumber\\
2\gamma f_{2}(x) = -2W(x) -(a+\epsilon +\gamma^{2}).
\label{fW}
\ea
Two cases have to be considered separately:
$ \gamma = 0 $ and $ \gamma \neq 0.$

\hspace*{3ex} For {\bf $ \gamma = 0 $} from (\ref{fW})
$ W(x) = -\frac{1}{2} (a+\epsilon)\equiv\kappa =\mbox{const}. $ Then from
Eqs.(\ref{1}) - (\ref{5}) one can derive that
\ba
f_{1}(x) = 3\kappa -f_{2}^{2}(x) + f_{2}^{\prime}(x);\nonumber\\
f_{0}(x) = 3\kappa f_{2}(x) -f_{2}^{3}(x) +f_{2}^{\prime\prime}(x)
-f_{2}(x)f_{2}^{\prime}(x);\nonumber\\
V(x) = f_{2}^{2}(x) - f_{2}^{\prime}(x) -2\kappa - a . \nonumber
\ea
Then the operator $Q^+$ is reduced ("modulo Hamiltonian $H$") to the {\bf first order}
operator
\ba
Q^+ = (\kappa - a - H)q^+.
\nonumber
\ea

\hspace*{3ex} More interesting case is for {\bf $ \gamma \neq 0 $}, when
$$ f_{2}(x) = -\gamma^{-1} W(x) - \nu;\quad \nu =\mbox{const}. $$
Substitution of this expression into Eq.(\ref{6}) gives equation,
which includes the function $W(x)$ only:
\be
(W^{\prime})^2 = \gamma^{-2}W^4 + 4(\nu\gamma^{-1}-1)W^3 + 4\nu^2W^2 + 4cW - 4d.
\label{new}
\ee
In general case its solution can be expressed in terms of elliptic functions,
but some particular values of constants (e.g. $\nu = c = d = 0$) in (\ref{new})
give simpler solutions as well.

It can be shown that now the operator $Q^+$ is expressed (again
"modulo Hamiltonian $H$") in terms of {\bf first order} and {\bf second order}
operators
\ba
Q^+ = L^+ - Hq^+;\quad L^+\equiv -\gamma^{-1}\partial^2 + (W-a)\partial - V\omega +
\omega^{\prime\prime}.
\nonumber
\ea
Therefore the same pair of Hamiltonians is intertwined both by the first order $q^+$
and the second order $L^+$ operators.
This result is in accordance with the general statements of the recent
paper \cite{andrianov} concerning "the
optimal set of two basic SUSY generators of even and odd order".
One can check also from the intertwining relations
that the third order operator $R\equiv (L^+)^{\dagger}q^+$ commutes with $H,$ playing
the role of its symmetry operator (see discussion in the preprint version of \cite{shape}
and \cite{andrianov}).

\vspace*{0.2cm}
\subsection*{\bf 5.4. \quad The classical limit}
\vspace*{0.2cm}
\hspace*{3ex} It is known that sometimes results obtained for
the quantum (supersymmetrical) systems can be successfully
used to investigate its classical prototype (e.g. see \cite{classic}
on two-dimensional second order SUSY
Classical Mechanics). We will consider here the
consequences of third order SUSY QM for the classical limit
$ \hbar \rightarrow 0. $ With this object we have to restore
the $ \hbar $ in all formulas of Section 2. It can be done
by rewriting differential operators in terms of the momentum
$p=-i\hbar\partial .$ The mnemonic rule is to multiply
each derivative by the multiplier $ \hbar $ in Eqs.(\ref{third}) -
(\ref{5}). Therefore, in the classical limit $ \hbar \rightarrow 0$
one obtains from (\ref{1}) - (\ref{4}):
\ba
\widetilde V - V &=& 0; \nonumber\\
(f_2^2-2f_1-3V)'&=&0;\nonumber\\
f_2'f_1-f_0'-f_2V'&=&0;\label{c1}\\
f_1V'-2f_0f_2'&=&0.\nonumber
\ea
The general solution of this "classical system" of equations
can be obtained from its quantum counterpart of Section 2 by
mnemonic rules formulated above.
It is useful to compare this problem with discussion in the
paper \cite{plyush1}, where the function
$ B(x, ip)=\sum b_{3-n}(ip)^n $ plays the role of classical
analogue of the quantum operator $ Q^{+} $  with
$b_{3-n}(x) \equiv f_n(x).$ These coefficient functions
have to satisfy the differential equation (Eq.(2.24)
of \cite{plyush1}):
\be
b_n'(x)+(5-n)b_{n-2}(x)V'(x)-L(x)b_{n-1}(x)=0, \label{c2}
\ee
which was not solved in the framework of pure classical theory
(see details in \cite{plyush1}). However Eq.(\ref{c2}) coincides
exactly (up to some normalization factors) with Eq.(\ref{c1}),
whose {\bf general solution} can be found analytically
from the quantum solution of Section 2 by $ \hbar\rightarrow 0$ limit.

\vspace*{0.1cm}
\hspace*{3ex}
\section*{\bf Acknowledgements}
Authors are grateful to A.A.Andrianov for useful discussions in the early
stage of the work and for reading of manuscript.
This work was partially supported by the Russian Foundation for Fundamental
Research (Grant No.02-01-00499).

\end{document}